\documentclass[10pt,a4paper,twoside]{article}
\usepackage{epsfig}
\usepackage{baltlat6}
\usepackage{array}
\begin{document}
\ \
\vspace{0.5mm}
\setcounter{page}{1}
\vspace{8mm}

\titlehead{Baltic Astronomy, vol.\,xx, yy, 2011}

\titleb{High energy emission of symbiotic
recurrent novae: RS Ophiuchi and V407 Cygni}

\begin{authorl}
\authorb{Margarita Hernanz}{1} and
\authorb{Vincent Tatischeff}{2}
\end{authorl}

\begin{addressl}
\addressb{1}{
Institut de Ci\`encies de l'Espai (CSIC-IEEC)\\
Campus UAB, Facultat de Ci\`encies, C5 par 2n, 08193 Bellaterra, Spain\\
hernanz@ieec.uab.es}

\addressb{2}{
Centre de Spectrom\'etrie Nucl\'eaire et de Spectrom\'etrie de Masse\\
CNRS/IN2P3 and Universit\'e Paris-Sud, 91405 Orsay, France\\
Vincent.Tatischeff@csnsm.in2p3.fr}

\end{addressl}

\submitb{}

\begin{summary} 
Recurrent novae occurring in symbiotic binaries are candidate sources of high energy photons, 
reaching GeV energies. Such emission is a consequence of particle acceleration leading to pion production. 
The shock between matter ejected by the white dwarf, undergoing a nova explosion, and the wind from the 
red giant companion is responsible for such a process, which mimics a supernova remnant but with much 
smaller energetic output and much shorter time scales. Inverse Compton can also be responsible for 
high energy emission. Recent examples are V407 Cyg, detected by Fermi, and RS Oph, which unfortunately 
exploded in 2006, before Fermi was launched.
\end{summary}

\begin{keywords} white dwarfs, novae, symbiotic stars, supernovae, X- and gamma-ray astronomy \end{keywords}

\resthead{
High energy emission of symbiotic recurrent novae}
{M. Hernanz \& V. Tatischeff}

\sectionb{1}{INTRODUCTION}

White dwarfs are the endpoints of stellar evolution of stars with masses smaller than about 10~$M_\odot$. In white dwarf stars there is not nuclear energy available anymore. Their chemical composition is either a CO or ONe mixture (it can also be pure He). Their typical masses are $\sim 0.6~M_\odot$ and their maximum mass is the Chandrasekhar mass ($\sim1.4~M_\odot$). White dwarfs cool down to very low luminosities (typically $L \sim10^{-4.5}~L_\odot$) when isolated, whereas they can explode when they are accreting matter in close binary systems.

There are two main types of binary systems where white dwarfs can accrete matter and subsequently explode. The most common case is the so-called cataclysmic variable, where the companion is a main sequence star. In this system, the mass transfer occurs via Roche lobe overflow, typical orbital periods are ranging from hours to days, and orbital separations are of several $10^{10}$~cm. Hydrogen burning in degenerate conditions on top of the white dwarf leads to a thermonuclear runaway and a (classical) nova explosion. Masses $\sim 10^{-5}$--$10^{-4}~M_\odot$ are ejected at velocities of several 100's or even 1000's of km~s$^{-1}$. A nova explosion does not disrupt the white dwarf (as occurs in thermonuclear - or type Ia - supernova explosions); therefore, after enough mass is accreted again from the companion star, a new explosion will occur. The typical recurrence time is $10^4$--$10^5$ years. In our Galaxy, about 35 classical nova explosions occur every year. 

Another scenario is the so-called symbiotic binary, where the white dwarf accretes matter from the stellar wind of a red giant companion. Typical orbital periods for these systems are a few 100 days (i.e. larger than in cataclysmic variables) and orbital separations are $10^{13}$--$10^{14}$~cm. This scenario leads to more frequent nova explosions than in cataclysmic variables, with typical recurrence periods smaller than 100 years. There are 10 known recurrent novae for which more than one outburst has been recorded and four of them are hosted by binaries with a red giant companion (Schaeffer 2010). 

\sectionb{2}{THE INTERESTING CASE OF RS OPH}

\begin{figure}
\begin{center}
\includegraphics[width=.6\textwidth]{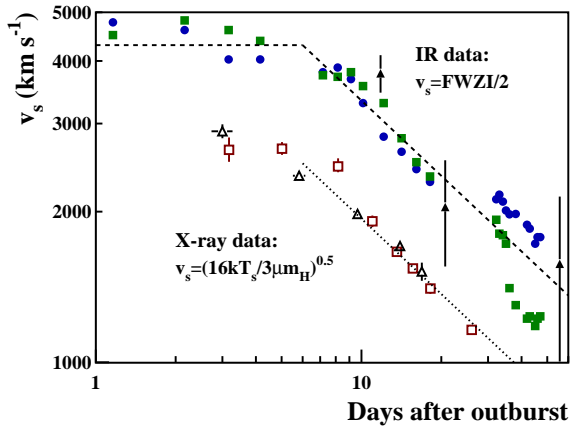}
\caption{Time-dependence of forward shock velocity in the 2006 outburst of RS Oph as deduced from
FWZI of IR emission lines (filled squares and circles: Das et al. 2006, filled triangles:
Evans et al. 2007) and X-ray measurements of the post shock temperature
(open triangles: Sokoloski et al. 2006, open squares: Bode et al. 2006). The
IR data can be modeled by (dashed line) $v_s (t) = 4300 (t / t_1)^{\alpha_v}$~km~s$^{-1}$, 
where $t_1 = 6$ days and $\alpha_v = 0 (-0.5)$ for $t  \le t_1 (t  >  t_1)$.}
\label{fig1}
\end{center}
\end{figure}

\begin{figure}
\begin{center}
\includegraphics[width=.8\textwidth]{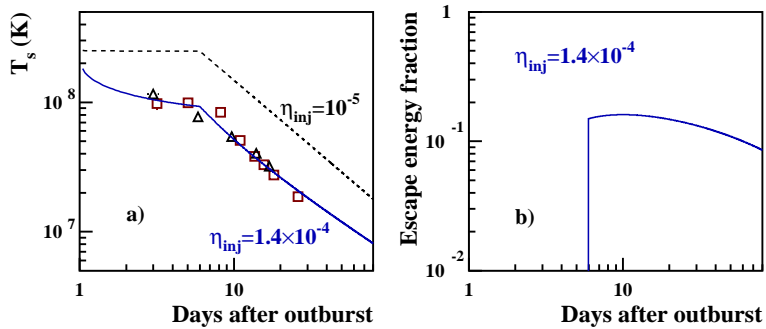}
\caption{a) Calculated postshock temperature for two values of $\eta_{\rm inj}$ compared to the RXTE and Swift data (same symbols as in Fig.~1). b) Cosmic-ray escape energy fraction, starting on day 6, for the best-fit acceleration efficiency $\eta_{\rm inj}$.}
\label{fig2}
\end{center}
\end{figure}

The recurrent nova of the symbiotic class RS Ophiuchi had its two last eruptions in 1985 and 2006. Its orbital period is 456~days, the recurrence period is 21~years, the estimated distance is $1.6\pm0.3$~kpc, and $\rm t_3=9.5$~days. 

A first issue is to find a self consistent model for the outburst of this nova, because it is not easy to find a combination of initial conditions leading to a thermonuclear runaway with such a short recurrence period. An extremely large initial mass and a high accretion rate are required (see Hernanz \& Jos\'e 2008). Initial white dwarf mass should be larger than $1.35~M_\odot$ and the mass accretion rate not smaller than $2\times 10^{-7}~M_\odot$~yr$^{-1}$ (see, e.g., Walder et al. 2008 and Worters et al. 2007, for some estimates of the expected accretion rate). 

Once the nova explodes, an expanding shock wave forms which sweeps the red giant wind: the system behaves as a ``miniature" supernova remnant, evolving much faster and being much dimmer. The characteristic timescale of evolution, which varies as $M_{\rm ej}^{3/2} v_{\rm wind}/[E_{\rm out}^{1/2}(dM/dt)_{\rm wind}]$, is much shorter than in the supernova case. Typical values for RS Oph are $M_{\rm ej}=3\times 10^{-6}~M_\odot$ for the ejected mass, $E_{\rm out}=10^{44}$ erg for the outburst energy, and $(dM/dt)_{\rm wind}=10^{-6}~M_\odot$~yr$^{-1}$ and $v_{\rm wind}=10$~km~s$^{-1}$ for the red giant wind mass loss rate and terminal velocity, respectively. Typical values for core collapse supernovae of type II (SN~II) are $M_{\rm ej}=10~M_\odot$, $E_{\rm out}=10^{51}$~erg, $(dM/dt)_{\rm wind}=10^{-5}~M_\odot$~yr$^{-1}$, and $v_{\rm wind}=10$~km~s$^{-1}$. Therefore, the characteristic time of evolution for RS Oph is about $10^{-5}$ times that for SN~II, such that the free expansion phase lasts just a few days in the nova remnant. 

The blast wave evolution of RS Oph during its 2006 outburst is shown in Figure 1 (see Tatischeff \& Hernanz 2007). The time dependence of the forward shock velocity as deduced from IR spectroscopic observations is compared to that from the X-ray observations with RXTE; the usual relation for a test-particle strong shock $v_s = [(16/3) (k T_s)/(\mu m_{\rm H})]^{1/2}$ was used to get shock velocities from X-ray fits of the postshock temperature $T_s$ (here $k$ is the Boltzmann constant and $\mu m_{\rm H}$ the mean particle mass). Two questions arise from a look at Figure 1: a) what makes the X-ray measurements of $v_s$ smaller than those from IR data? b) Why did the cooling phase started as early as 6 days, when $T_s$ was about $10^8$~K and thus radiative cooling was not important? The answer to both question is cosmic rays. 

Acceleration of cosmic rays can influence the evolution of the nova remnant, mainly because of the energy loss due to the escape of the highest energy particles from the blast wave region. Then, a good agreement between the IR and X-ray measurements of the shock temperature can be obtained (see Figure 2a) with a moderate acceleration efficiency $\eta_{\rm inj} \sim 10^{-4}$ (Tatischeff \& Hernanz 2007), this parameter being the fraction of total shocked protons in protons injected from the thermal pool into the diffusive acceleration process. The fraction of the total energy flux processed by the shock that escaped via diffusion away from the shock system of the highest energy particles is shown in Figure 2b. The corresponding energy loss rate is estimated to be $\sim 2 \times 10^{38} (t/{\rm 6~days})$~erg~s$^{-1}$ (Tatischeff \& Hernanz 2007), which is approximately 100 times the bolometric luminosity of the postshock plasma 6 days after outburst. This means that energy loss via escape of accelerated particles is much more efficient than radiative losses to cool the shock.

A prediction of the $\gamma$-ray emission associated to the accelerated particles has been made. The neutral pion ($\pi^0$) production has been calculated from the gas density in the red giant wind (i.e. $(dM/dt)_{\rm wind}$ and $v_{\rm wind}$) and the cosmic-ray energy density required to explain the IR and X-ray observations. The Inverse Compton (IC) contribution has been estimated from the nonthermal synchrotron luminosity $L_{\rm syn}$ and the ejecta luminosity $L_{\rm ej}$: $L_{\rm IC}=L_{\rm syn} \times U_{\rm rad}/(B^2/8 \pi)$, where $U_{\rm rad} \sim L_{\rm ej}/(4\pi cR_s^2)$ ($R_s$ is the shock radius and $c$ the speed of light) and $B$ is the postshock magnetic field. According to early radio detections of RS Oph at frequencies lower than 1.4 GHz by Kantharia et al. (2007): $L_{\rm syn} \sim 5 \times 10^{33} (t/{\rm 1~day})^{-1.3}$~erg~s$^{-1}$. The ejecta luminosity can be assumed to be equal to the Eddington luminosity in novae in outburst (i.e. approximately $10^{38}$~erg~s$^{-1}$). We then find $L_{IC} \sim L_{\rm syn}$ and can conclude that $\gamma$-rays come mainly from $\pi^0$ production (see Figure 3).

Finally, predicted light curves for $\gamma$-ray energies $E_\gamma > 100$~MeV and $E_\gamma > 30$~GeV are compared in Figure 4 to the Fermi/LAT sensitivities for 1 week and 1 month observation times. The conclusion is that RS Oph (2006) would have been detected by Fermi/LAT. The origin of the high energy $\gamma$-rays is the collision of the nova blast wave with the dense wind of the red giant companion; such a situation can only be found in novae occurring in symbiotic binaries. 

\begin{figure}
\begin{center}
\includegraphics[width=.6\textwidth]{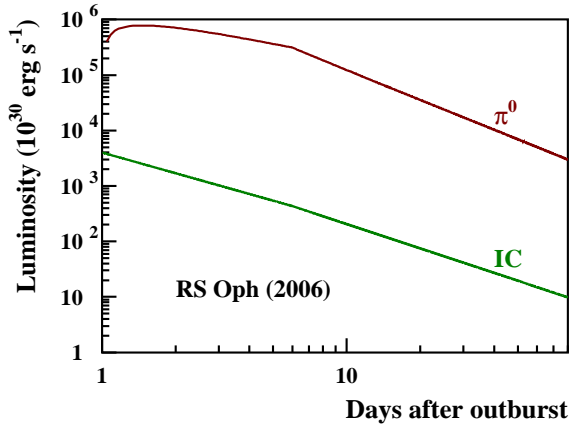}
\caption{Evolution with time from outburst of the $\gamma$-ray luminosity (in units of $10^{30}$erg~s$^{-1}$) corresponding to pion production and IC emission in RS Oph (2006).}
\label{fig3}
\end{center}
\end{figure}

\begin{figure}
\begin{center}
\includegraphics[width=.62\textwidth]{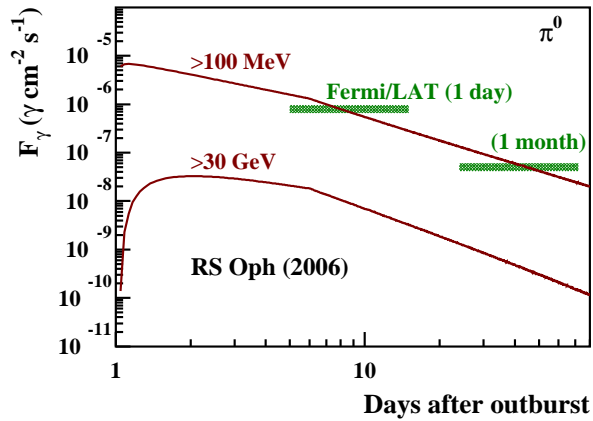}
\caption{High energy $\gamma$-ray light curves expected for RS Oph (2006), as compared with Fermi/LAT sensitivities for 
$E_\gamma > 100$~MeV.}
\label{fig4}
\end{center}
\end{figure}

\sectionb{3}{V407 CYG: THE FIRST NOVA DETECTED IN HIGH ENERGY $\gamma$-RAYS}

V407 Cygni is a symbiotic Mira binary system, composed of a white dwarf and an asymptotic giant branch (AGB) star of the Mira type. The pulsation period of the Mira is approximately 750~days and the orbital period of the system is $\sim$43~years (Munari et al. 1990). V407 Cyg underwent an outburst on 10 March 2010, with a spectroscopic evolution resembling that of the 2006 outburst of RS Oph (Shore et al. 2011). Fermi/LAT detected it on the same day and then repeatedly for a few days, with a peak flux in $\gamma$-rays observed between 13 to 14 March 2010, i.e. about 3--4 days after the optical outburst (Abdo et al. 2010). So, our predictions for RS Oph can now be tested with real observations with Fermi/LAT. However, less multiwavelength observations are available for V407 Cyg (2010) than for RS Oph (2006). 

There are two main differences between the two objects. First, V407 Cyg is not a standard recurrent nova, as RS Oph, since no regular eruptions were known before 2010 (Munari et al. 2011). Second, the orbital period is much larger for V407 Cyg than for RS Oph (43 years vs. 456 days); then, the separation between the white dwarf and the AGB star in V407 Cyg is very large (about 15~AU, 10 times larger than in RS Oph). This means that the shock wave needs about 7 days to reach a distance equal to the binary separation, and thus it should propagate through the red giant wind perturbed by the orbital motion. On the contrary, in RS Oph free expansion of the shock wave through the unperturbed wind occurred from day one after outburst. We also note that nonthermal radio synchrotron emission is not (yet) reported for V407 Cyg (2010), contrary to RS Oph (2006).

We have made a preliminary estimate of the $\gamma$-ray flux from neutral pion decay in V407 Cyg (2010), getting $F_\gamma(E_\gamma > {\rm 100~MeV}) \sim 10^{-6}$~photons~cm$^{-2}$~s$^{-1}$, for an estimated mean postshock density for a few days after outburst of $\sim 2 \times 10^9$~cm$^{-3}$, a nova energy output of $\sim 10^{44}$~erg and a distance of 2.7 kpc. This result is consistent with the peak $\gamma$-ray flux detected by Fermi/LAT about 3--4 days after the optical outburst.

\sectionb{4}{SUMMARY AND DISCUSSION}

Recurrent novae in symbiotic binaries are expected to accelerate particles and emit high-energy $\gamma$-rays detectable with Fermi/LAT, because of the shock wave propagation in the dense wind expelled by the red giant star. According to our calculations, RS Oph (2006) would have been detected by Fermi. V407 Cyg (2010) was detected by Fermi, but did not behave as RS Oph regarding X-ray and radio emissions. Other similar systems exist in the Galaxy: eventually 1-2 symbiotic novae per year are expected (but not necessarily detected in the optical).


\References

\refb Abdo, A.~A., Ackermann, M., Ajello, M., et al.\ 2010, Science, 329, 817 

\refb Bode, M. F., et al. 2006, ApJ, 652, 629

\refb Das, R., Banerjee, D. P. K., \& Ashok, N. M. 2006, ApJ, 653, L141

\refb Evans, A., et al. 2007, MNRAS, 374, L1

\refb Hernanz, M., \& Jos{\'e}, J.\ 2008, New Astron. Rev., 52, 386 

\refb Kantharia, N.~G., Anupama, G.~C., Prabhu, T.~P., et al.\ 2007, ApJ, 667, L171 

\refb Munari, U., Margoni, R., \& Stagni, R.\ 1990, MNRAS, 242, 653 

\refb Munari, U., Joshi, V.~H., Ashok, N.~M., et al.\ 2011, MNRAS, 410, L52 

\refb Schaefer, B.~E.\ 2010, ApJS, 187, 275 

\refb Shore, S.~N., Wahlgren, G.~M., Augusteijn, T., et al.\ 2011, A\&A, 527, A98 

\refb Sokoloski, J. L., Luna, G. J. M., Mukai, K., \& Kenyon, S. J. 2006, Nature, 442, 276

\refb Tatischeff, V., \& Hernanz, M.\ 2007, ApJ, 663, L101 

\refb Walder, R., Folini, D., \& Shore, S.~N.\ 2008, A\&A, 484, L9 

\refb Worters, H.~L., Eyres, S.~P.~S., Bromage, G.~E., \& Osborne, J.~P.\ 2007, MNRAS, 379, 1557 

\end{document}